\documentclass[12pt]{article}
\begin{document}
\sf
\begin{center}
   \vskip 2em
{\LARGE \sf
Conformally invariant bending energy for hypersurfaces}
\vskip 3em
 {\large \sf  Jemal Guven \\[2em]}
\em{
Instituto de Ciencias Nucleares,
 Universidad Nacional Aut\'onoma de M\'exico\\
 Apdo. Postal 70-543, 04510 M\'exico, DF, MEXICO\\[1em]
}
\end{center}
 \vskip 1em

\begin{abstract}
The most general conformally invariant 
bending energy of a closed four-dimensional surface, 
polynomial in the extrinsic curvature and its derivatives, is constructed. This invariance manifests itself
as a set of constraints on the corresponding stress tensor.
If the topology is fixed, there are three independent polynomial invariants: 
two of these are the straighforward
quartic analogues of the quadratic Willmore energy for a two-dimensional surface;
one is intrinsic (the Weyl invariant), the other
extrinsic; the third invariant involves a sum of a quadratic in 
gradients of the extrinsic curvature --- which is not itself invariant --- and a  
quartic in the curvature. 
The four-dimensional energy quadratic in extrinsic curvature 
plays a central role in this construction.
\end{abstract}
\today \vskip 1em PACS: 04.60.Ds,  
87.16.Dg,
46.70.Hg,
02.40.Hw
\vskip 3em
\section{\sf Introduction}

The bending energy of a two-dimensional surface, quadratic in its
extrinsic curvature, is invariant under scaling; size does not matter. 
What is less obvious is that this energy is also invariant under
transformations of the three-dimensional background which
preserve angles; it is conformally invariant. 
Any two surface geometries related to each other by 
inversion in a point have the same energy. This property was first
studied systematically by Willmore in the 60s \cite{Willmore}. More recently,  
it was discovered to lie at the heart of some fascinating connections between
differential geometry and integrable systems \cite{Konop}. 
At mesoscopic scales, the physics of a fluid membrane ---
formed by the spontaneous aggregration of
amphiphilic molecules into bilayers in water --- is
captured completely by its geometrical degrees of freedom
\cite{Nel.Pir.Wei:89,LesHouches}. On such
scales the membrane can be modeled as a two-dimensional surface; at lowest order,
the principal cost in energy is associated with bending this surface \cite{CanHel}. Remarkably, 
all of the molecular details get telescoped into a single 
rigidity modulus.  A role was also found for a relativistic counterpart in the eighties:
an addition quadratic in extrinsic curvature to the action of a
relativistic string accounts for the behaviour of colour 
flux tubes in QCD \cite{Polyakov,Polyakov1,Kleinert}.

Higher-dimensional analogues of the two-dimensional bending energy are also of potential 
interest both as statistical field theories, and relativistically, as braneworld actions. 
If the surface dimension differs from two, however, the 
energy quadratic in extrinsic curvature is no longer scale invariant
much less conformally invariant: a higher-dimensional sphere without a constraint on its
area will collapse; tension is necessarily introduced. 
Conformal invariant energies, polynomial in the 
extrinsic curvature, are however simple to construct: the building block 
is the traceless part of the 
extrinsic curvature tensor which transforms by a multiplicative factor\cite{Willmore,Willmore99}. 
In four dimensions, 
for a hypersurface of fixed topology, there are two independent conformal invariant energies 
quartic in the extrinsic curvature.The snag is 
that these invariants alone
cannot accurately describe a conformally invariant theory of bending. 
The reason is simple: when expanded as a power series in terms of a
height function, they begin with a term quartic in this function.
Thus, they  vanish in the Gaussian approximation to the
energy truncating it at the quadratic in the height function. 
In particular, there is no harmonic regime
to describe fluctuations about a flat geometry. On dimensional grounds, the 
relevant invariant must involve 
curvature gradients. This invariant will play a role 
in the formulation of a consistent statistical field 
theory of four-dimensional hypersurfaces.
Its identification, however, is  somewhat
less trivial than that of its quartic counterparts.
To do this, it will  be useful to approach the problem from a global point of view
which focuses directly on the transformation properties of the surface energy
rather than the individual tensors which appear within it.
For the sake of simplicity, we will focus on a closed hypersurface of fixed topology;
more simple still, think topological four-sphere.

Consider any energy, constructed using the metric and the extrinsic 
curvature, which is invariant under reparametrizations of the hypersurface and 
Euclidean motions of space. 
If the theory is invariant under translations the response of this energy  
to an arbitrary infinitesimal deformation of the hypersurface can be 
expressed in terms of a stress tensor \cite{Stress,auxil}. 
In \cite{auxil}
it was shown that the tangential stress $f^{ab}$ has two contributions:
one of these is the metric stress tensor $T^{ab}$ which 
determines the response of the energy to changes in the 
intrinsic geometry; the second,
which determines the response to changes in the extrinsic geometry, involves the 
functional derivative of the energy with respect to the extrinsic curvature,
${\cal H}^{ab}$. 
One is interested, in particular, in determining the response of the energy
to the deformation of the hypersurface induced by an infinitesimal conformal transformation.
It is possible to characterize this response in a 
remarkably succinct way in terms of traces: that of $f^{ab}$ and that of ${\cal H}^{ab}$.
Conformal invariance will place constraints on these traces. 
In contrast to a conformal invariant of the intrinsic geometry which has a representation with
${\cal H}^{ab}=0$, these constraints may be satisfied in a very subtle way by an invariant of the 
the extrinsic geometry.

The task is to identify energies that are consistent with these constraints. 
While the focus will be on closed four-dimensional hypersurfaces,
the techniques developed will be independent of the dimension. 
We first briefly describe the construction, within this framework,
of the two well-known four-dimensional conformally invariant 
energies quartic  in extrinsic curvature. 
Modulo the Gauss-Codazzi equations,
which identify the intrinsic Riemann tensor with 
a quadratic in the extrinsic curvature, one of these invariants is 
the Weyl invariant associated with the intrinsic geometry of the hypersurface, 
and thus insensitive to the particular way the hypersurface is embedded. 
The third invariant 
involves a balance of a part quadratic in curvature gradients
with a quartic in curvature;
neither term on its own is conformally invariant.
We show how this constraint can be satisfied by tuning the quartic 
so that the two trace terms cancel. Intriguingly, this cancellation 
involves properties of the four-dimensional Willmore energy quadratic in the
extrinsic curvature 
(which is not itself a conformal invariant in this dimension) in an essential way.

\section{\sf Linear response, the Euler-Lagrange derivative as a divergence,
and the stress}

Consider a closed $D$-dimensional hypersurface
embedded in $R^{D+1}$. This hypersurface is described locally by the
embedding, ${\bf x}= {\bf X} (\xi^a)$.
Here ${\bf x}=(x^1,\cdots,x^{D+1})$ and $\xi^a$, $a=1,\dots,D$
parametrize the hypersurface. The metric tensor and extrinsic curvature
induced by ${\bf X}$ are respectively $g_{ab}= {\bf
e}_a\cdot {\bf e}_{b}$ and $K_{ab}= {\bf e}_a\cdot \partial_b {\bf
n}$, where ${\bf e}_a= \partial_a {\bf X}$, $a=1,\dots,D$ are
tangent and ${\bf n}$ is the unit normal. The
Gauss-Weingarten equations are $\nabla_a {\bf
e}_b = - K_{ab} {\bf n}$ and $\partial_a {\bf n} = K_a{}^b {\bf
e}_b$ \cite{Spivak}. $\nabla_a$ is the covariant derivative compatible with $g_{ab}$;
spatial indices get raised with the inverse metric $g^{ab}$.
We are interested in functionals of ${\bf X}$ 
which are invariant under reparametrizations of the hypersurface.

The metric and extrinsic curvature are both invariant under
the change in ${\bf X}$ induced by a Euclidean motion in $R^{D+1}$:
a Euclidean invariant energy $H[{\bf
X}]$ can therefore be cast as a functional of the metric, the extrinsic
curvature and its derivatives,
\begin{equation}
H [{\bf X}]= \int dA {\cal H}(g_{ab}, K_{ab},\nabla_a
K_{bc},\cdots) \,.\label{eq:Hdef}
\end{equation}
The area element induced on the
hypersurface is $dA =\sqrt{{\rm det} g_{ab}} d^D\xi$.
We wish, in particular, to construct an  energy
which is invariant
under deformations induced by a conformal change of the Euclidean background:
\begin{equation}
\delta {\bf x}=
{\bf a} + {\bf B} {\bf x} + \lambda {\bf x} + 
 {\bf x}^2\, {\bf c} - 2 ({\bf c}\cdot {\bf x})\, {\bf x}\,,
\end{equation}
where 
${\bf a}$ and ${\bf c}$ are two constant vectors (${\bf c}$ has dimensions of inverse length),
${\bf B}$ is an antisymmetric $(D+1)\times (D+1)$ matrix, and $\lambda$ is a positive constant.
$\delta {\bf x}$
is the sum  of an infinitesimal Euclidean motion, a change of scale and a special 
conformal transformation. The latter exponentiates to 
the composition of an inversion ${\bf x}\to {\bf x}/
{\bf x}^2$, a translation through the vector ${\bf c}$ , and another
inversion, ${\bf x} \to 
({\bf x}+ {\bf x}^2 {\bf c})/ (
 1+ 2{\bf c}\cdot {\bf x} + {\bf c}^2 \,{\bf x}^2)$.
Both $g_{ab}$ and $K_{ab}$ are invariant under Euclidean 
motions: as a result, any energy of the form (\ref{eq:Hdef})
will also be by construction. What one now needs to do is 
characterize the constraints placed on $H$ by invariance under 
scaling and special conformal transformations. 
 
It is useful to first determine the linear response of
$H$ to any small deformation
of the hypersurface. This task is simplified by exploiting
invariance under Euclidean motions of the ambient space.
While Noether's theorem informs us that the Euler-Lagrange derivative
can always be cast as the divergence of a stress tensor,
in all but the simplest case --- an energy proportional to 
the area functional describing surface tension --- the
identification of this tensor is subtle: unlike the stress associated with
area, the stress will depend not only on the intrinsic geometry
but also on 
how the hypersurface bends. The tug on the hypersurface will possess a normal component.

A small deformation of the hypersurface is described by the
infinitesimal change in the embedding functions ${\bf X}$
\begin{equation}
{\bf X} \to {\bf X} + \delta {\bf X}\,.
\end{equation}
Note the following points:

\vskip1pc \noindent (1) as a consquence of the reparametrization invariance of $H$
in a closed geometry,
 the response of $H$ is independent of the
tangential projection of $\delta {\bf X}$; thus
\begin{equation}
\delta H = \int dA \, {\cal  E}\,{\bf n} \cdot\delta {\bf X}
\label{eq:delH0}
\end{equation}
involves only the normal projection. 
The Euler-Lagrange derivative of $H$ with respect to ${\bf
X}$ is denoted by ${\cal E}{\bf n}$.

\vskip1pc
\noindent (2) the translational invariance
of $H$ implies that its  Euler-Lagrange derivative is  a
divergence \cite{Stress,auxil}
\begin{equation}
{\cal  E} {\bf n} = \nabla_a {\bf f}^a \,. \label{eq:Ef}
\end{equation}
The hypersurface current ${\bf f}^a\cdot {\bf a}$ is
associated with the invariance of $H$ under a translation
$\delta {\bf X}= {\bf a}$. When the Euler-Lagrange equation 
${\cal E}=0$ is satisfied, this current is conserved.
The closure of the geometry then permits
$\delta H$ to be recast in the remarkably simple form
\begin{equation}
\delta H = -
 \int dA \, {\bf f}^a \cdot \nabla_a\delta {\bf X}\,.
\label{eq:delH02}
\end{equation}
This expression involves one less derivative than
Eq.(\ref{eq:delH0}). Note that one does not need to know
how ${\bf f}^a$ itself transforms. Eq.(\ref{eq:delH02}) is valid whether or not 
the Euler-Lagrange equation are satisfied.
This equation will be used to 
examine the response of $H$ to the deformation in the hypersurface
induced by conformal transformations of space.

\section{\sf The stress}

The stress ${\bf f}^a$ associated with $H$ is given by
\begin{equation}
{\bf f}^a = (T^{ab} - {\cal H}^{ac}\, K_{c}{}^b)\, {\bf e}_b -
\nabla_b {\cal H}^{ab} \,{\bf n}\,,
\label{eq:fabfa}
\end{equation}
where 
${\cal H}^{ab}$ is the functional
derivative of $H$ with respect to $K_{ab}$,
\begin{equation}
{\cal H}^{ab} = 
{\partial {\cal H}\over\partial K_{ab} }-
\nabla_c \, \left({\partial {\cal H}\over\partial \nabla_c K_{ab}}\right)
 + \cdots
\label{eq:calHdef}
\end{equation}
and $T^{ab} =- (2/\sqrt{g})\, \delta H/ \delta g_{ab}$
is the intrinsic stress
tensor associated with the metric $g_{ab}$. 
This construction involves treating $g_{ab}$ and $K_{ab}$ as
independent variables in ${\cal H}$; to do this consistently requires
one to  introduce  a set of 
auxiliary variables to constrain $g_{ab}$ and $K_{ab}$ to satisfy 
the Gauss-Weingarten structural relationships. 
The ellipsis appearing on the 
r.h.s of Eq.(\ref{eq:calHdef}) indicates terms which appear if
${\cal H}$ depends on derivatives of $K_{ab}$ higher than first.
A simple derivation of Eq.(\ref{eq:fabfa}) is provided in \cite{auxil}.
We note, in particular, that ${\bf f}^a$ decomposes into tangential and
normal parts:
\begin{equation}
{\bf f}^a = f^{ab}{\bf e}_b + f^a {\bf n}\,. \label{eq:fdecomp}
\end{equation}

\vskip1pc
\noindent This decomposition has the following properties which are relevant:

\vskip1pc \noindent (1) The tangential projections of
Eq.(\ref{eq:Ef}) provide a consistency condition on the
components of the stress
\begin{equation}
\nabla_a f^{ab} + K^{ab}f_a =0\,;
\label{eq:Eft}
\end{equation}
the normal component determines ${\cal E}$:
\begin{equation}
{\cal E}= \nabla_a f^a - K_{ab} f^{ab}\,. \label{eq:Efn}
\end{equation}
\vskip1pc\noindent
(2) In general, the normal stress $f^a$ is a divergence.

\vskip1pc \noindent (3) Even though both ${\cal H}^{ab}$ and
$T^{ab}$ are symmetric tensors, $f^{ab}$ will not generally be
symmetric. On one hand, as Eq.(\ref{eq:Efn}) indicates clearly, only
the symmetric part of $f^{ab}$ contributes to the Euler-Lagrange
derivative. An anti-symmetric contribution, if present, will however 
show up in the consistency conditions (\ref{eq:Eft}) and so cannot be discarded naively.

Let us now consider specific forms for the function
${\cal H}$ appearing  in Eq.(\ref{eq:Hdef}) which will be used in the 
construction of conformal invariants.

\subsection{\sf ${\cal H}(g_{ab},K_{ab})$}
\label{sec:gK}

Suppose that ${\cal H}$ does not involve derivatives of $K_{ab}$:
${\cal H}= {\cal H}(g_{ab}, K_{ab})$.
Then
\begin{equation}
{\cal H}^{ab} = \partial {\cal H}/ \partial K_{ab}\,,
\end{equation}
and it is simple to show that 
\begin{equation}
T^{ab}= 2 {\cal H}^{ac} K_c{}^b -
{\cal H} g^{ab}\,,
\end{equation}
so that
\begin{equation}
{\bf f}^a =
( {\cal H}^{ac}K_{c}{}^b -{\cal H} g^{ab}) 
{\bf e}_b - \nabla_b {\cal H}^{ab} {\bf n} \,.
\label{eq:fK}
\end{equation}
Note that $f^{ab}$ is manifestly symmetric.
A straighforward calculation gives
\begin{equation}
{\cal E}= -\nabla_a \nabla_b {\cal H}^{ab} - K_{ac} K^c{}_b {\cal
H}^{ab} + {\cal H}K\,,
\end{equation}
where $K= g^{ab}K_{ab}$ is the trace of $K_{ab}$ ($D$ times the
mean curvature).

 \vskip1pc\noindent In particular, for the Canham-Helfrich or Willmore energy, 
one has
\begin{equation}
H_0={1\over 2}\int dA \, K^2 \,, 
\label{eq:H0}\end{equation} 
${\cal H}= K^2/2$ and ${\cal
H}^{ab} = K g^{ab}$, so that
\begin{equation}
{\bf f}^a=  K (K^{ab}- {1\over 2}g^{ab} K)
 \,{\bf e}_b  - \nabla_a K {\bf n}\,,
\label{eq:fhel}
\end{equation}
and \cite{Willmore} (see also 
\cite{Hel.OuY:87})
\begin{equation}
{\cal E}= -\, \nabla^2 K + {1\over 2} K (K^2 - 2\, K^{ab} K_{ab}) \,,
\label{eq:Ehel}
\end{equation}
If $K=0$, then ${\cal E}=0$.
The relationship between ${\cal E}$
and ${\bf f}^a$ for $H_0$ will play a role in
the construction of a higher-derivative conformal invariant of a
four dimensional hypersurface.

Note that it is unnecessary to admit an
explicit intrinsic curvature dependence in ${\cal H}$. This is
because the Gauss-Codazzi equations \cite{Spivak}
\begin{equation}
{\cal R}_{abcd}= K_{ac} K_{bd} - K_{ad} K_{bc} \label{eq:GC}
\end{equation}
completely fix the Riemann tensor, as well as its contractions,
the Ricci tensor ${\cal R}_{ab}= g^{cd} R_{acbd}$ and the scalar
curvature ${\cal R}= g^{ab} {\cal R}_{ab}$, in terms of the
extrinsic curvature.\footnote{\sf The Riemann tensor is defined 
intrinsically by the failure of the $\nabla_a$ to commute: for a space vector $V_a$,
we have the Ricci identity
$(\nabla_a\nabla_b -\nabla_b \nabla_a) V_c = {\cal R}_{abc}{}^d V_d$.}
 However, if one is interested explicitly in
a functional of the intrinsic geometry, ${\cal H}= {\cal
H}(g_{ab}, {\cal R}_{abcd}, \nabla_e {\cal R}_{abcd}, \dots)$, it
may then be more appropriate to treat these tensors as
functionals of $g_{ab}$ alone, and ignore the integrability
conditions (\ref{eq:GC}). If this is done, ${\cal H}^{ab}=0$
and $T^{ab}$ is the
stress tensor of the (purely) metric theory defined by ${\cal H}$.
Now, $f^{ab}=T^{ab}$ and it is manifestly symmetric; furthermore
$f^a=0$.\footnote {\sf Intrinsically defined invariants are not the
only ones possessing this property. One can show that the
geometrical invariants constructed out of the
symmetric polynomials in the curvature,
$P_N(\sigma_1,\cdots,\sigma_D)$,  $N\le D$, where $\{\sigma_i\}$
are the principal curvatures, do also. For example, when $N=1$,
$P_1= K$ and $f^{ab} = K^{ab} - g^{ab} K$, which is conserved, so
that $f^a=0$.} The Euler-Lagrange derivative is then simply ${\cal
E}= - K_{ab} T^{ab}$; the consistency condition then reads
$\nabla_a T^{ab}=0$ --- the metric stress tensor is conserved.
Clearly, it does not matter how one decides
to split the burden on $g_{ab}$ and $K_{ab}$, so long as it is done 
consistently when performing the variations in the derivation of
${\bf f}^a$. As discussed in detail elsewhere, if  ${\bf f}^a$ is treated
as a differential form, the difference between its value in the two representations
is an exact form.

\subsection{\sf ${\cal H}(g_{ab},K_{ab},\nabla_c K_{ab})$}

If one extends the
class of functionals to include a dependence on
$\nabla_c K_{ab}$, there are few useful
general statements concerning the structure of
${\bf f}^a$.
Our limited goal, however, is to identify conformal invariants
of closed hypersurfaces so we do not need to consider the most general form.

Consider candidate polynomials in $\nabla_a K_{bc}$ and $K_{ab}$
that are consistent with scale invariance.
When $D=2$, there are none. When $D=3$, there is 
a Chern-Simons type topological energy; it vanishes on a closed 
geometry. When $D=4$, the
quadratics in $\nabla_c K_{ab}$ are scale invariant. 
As shown in \cite{defos}, however, any quadratic in derivatives of $K_{ab}$
is expressible, modulo a divergence, as a
sum of the simple invariant
\begin{equation}
H_1= {1\over 2}\int dA\, (\nabla K)^2 \label{eq:nabK2}
\label{eq:H1def}
\end{equation}
and an integral over some quartic in $K_{ab}$. The latter
is of the form, ${\cal H}(g_{ab},K_{ab})$,
already considered in section (\ref{sec:gK}).
So $H_1$ is the only  invariant that needs to be considered.

The demonstration of this claim involves the
Codazzi-Mainardi integrability
conditions, 
\begin{equation}
\nabla_a K_{bc}- \nabla_b K_{ac}=0\,,
\label{eq:cm}
\end{equation}
as well as the Ricci identity applied to $K_{ab}$
\begin{equation}
[\nabla_a,\nabla_b] K_{cd} =
R_{abc}{}^f K_{fd} + R_{abd}{}^f K_{cf}\,.
\label{eq:Riccid}
\end{equation}
Consider the energy constructed using the quadratic
$\nabla_a K_{bc} \nabla^a K^{bc}$.
One first uses Eq.(\ref{eq:cm}) followed by an integration by
parts to obtain
\begin{equation}
\int dA\; (\nabla_a K_{bc}) (\nabla^a K^{bc} ) = \int dA \;
(\nabla_a K_{bc})(\nabla^b K^{ac}) = - \int dA \; K^b{}_{c}
\nabla_a\nabla_b K^{ac} \,.
\end{equation}
One then makes use of Eq.(\ref{eq:Riccid}) to switch derivatives
so that
\begin{eqnarray}
\int dA (\nabla_a K_{bc}) (\nabla^a K^{bc} ) &=& -
\int dA \Big(
K_{c}{}^b
\nabla_b \nabla_a K^{ac} -
{\cal R}_{abcd} K^{ac} K^{bd} +
{\cal R}_{ab} K^{ac}K_c{}^b\Big)\,.
 \end{eqnarray}
The contracted Codazzi-Mainardi equations, $\nabla_a K^{ab}-
\nabla^b K=0$, and another integration by parts are applied to
the first term to nudge it into the required form:
\begin{equation}
\int dA
K_{c}{}^b
\nabla_b \nabla_a K^{ac} =
 - \int dA \,\ \nabla_b K^{b}{}_{c} \nabla_a K^{ac}
= - \int dA \,\ \nabla^c K \nabla_c K\,.
\end{equation}
One concludes that
\begin{eqnarray}
\int dA \; (\nabla_a K_{bc}) (\nabla^a K^{bc} ) &=&
\int dA \,\Big( \nabla^c K \nabla_c K
+  {\cal R}_{abcd} K^{ac} K^{bd} -
{\cal R}_{ab} K^{ac}K_c{}^b
\Big)\nonumber\\
&=& \int dA \,\Big( \nabla^c K \nabla_c K
+  ({\rm tr} \, K^2)^2 - K {\rm tr}\, K^3
\Big)\,.
\label{eq:bef}
\end{eqnarray}
The notation
${\rm tr}\, K^n
= K_{a_1}{}^{a_2}\cdots K_{a_n}{}^{a_1}$ has been introduced.
On the second line, the Gauss-Codazzi equations (\ref{eq:GC}) have been used to
eliminate the Riemann tensor in favour of a quadratic in extrinsic curvature.
The energy  $H_1$, given by Eq.(\ref{eq:nabK2}), is reproduced 
modulos a quartic in extrinsic curvature. 

Note that for $H_1$ one has (this is true for any dimension $D$),
\begin{equation}
{\cal H}^{ab}= -\nabla_c (g^{ab}\nabla^c K)
=- g^{ab}\nabla^2 K\,,
\end{equation}
and
\begin{equation}
T^{ab}= \nabla^a K\nabla^b K
 - {1\over 2} g^{ab} (\nabla K)^2
- 2 K^{ab}  \nabla^2 K \,.
\end{equation}
The second derivative term originates in the variation of
$\nabla_a K$ with respect to $g_{ab}$. The correponding stress
tensor is \cite{MDG}
\begin{equation}
{\bf f}^a = \left[\nabla^a K \nabla^b K  - {1\over 2} g^{ab} (\nabla K)^2
- K^{ab} \nabla^2 K\,\right]\, {\bf e}_b
+  \nabla^a \nabla^2 K\, {\bf n}\,.
\label{eq:fnabK}
\end{equation}
Again $f^{ab}$ is symmetric.\footnote {\sf Evidently, one has to proceed to a 
relatively high
order energy to produce an $f^{ab}$ which is not symmetric: for
${\cal H}= K^{ab}\nabla_a K \nabla_b K$, ${\cal H}^{ab}$ does not 
commute with $K_{ab}$ and thus $f^{ab}$ is not symmetric.}  
In this case, it is simple to check
that
\begin{equation}
{\cal E}= [\nabla^2 + {\rm tr}\, K^2] \, \nabla^2 K -
 K^{ab} [\nabla_a K \nabla_b K  - {1\over 2} g_{ab} (\nabla K)^2] \,.
\end{equation}
Note that if $K=0$, then ${\cal E}=0$ so that minimal hypersurfaces also
minimize $H_1$.

\section{\sf Scaling}

On one hand, it is trivial to identify energies which are scale invariant.
On the other, the imprint of scale invariance on the stress tensor is subtle 
and it will be relevant to our interpretation of the response, under special conformal transformations, 
of the energy in terms of the stress tensor.

Consider an energy with a fixed scaling dimension.
Under a change of scale ${\bf X} \to \Lambda {\bf X}$, where $\Lambda$ is a positive 
constant, one has
\begin{equation}
H[\Lambda {\bf X}]= \Lambda^{D+d}H[{\bf X}]
\label{eq:Hlam}
\end{equation}
for some 
$d$, or alternatively, in terms of the corresponding density,
${\cal H} [\Lambda {\bf X}]= \Lambda^d
{\cal H} [{\bf X}]$.
$H$ is scale invariant when $d=-D$.

Consider now an infinitesimal change of scale,
$\Lambda=1+\lambda$;
at first order in $\lambda$, Eq.(\ref{eq:Hlam}) gives
\begin{equation}
\delta_\lambda H = (D+ d) \lambda H \,.
\label{eq:sinf}
\end{equation}
On the other hand, the first order variation
Eq.(\ref{eq:delH02}) with the substitution 
$\delta{\bf X}= \lambda {\bf X}$ 
expresses $\delta_\lambda H$ in terms of the trace of the tangential stress,
\begin{equation}
\delta_\lambda H =
- \lambda \int dA \, f^a{}_a \,,
\label{eq:delHs}
\end{equation}
where $f^a{}_a= g_{ab} f^{ab}$. 
Comparison of Eq.(\ref{eq:sinf}) with Eq.(\ref{eq:delHs})
furnishes an identity,
\begin{equation}
(D+ d)  H = - \int dA \, f^a{}_a \,.
\label{eq:delHS}
\end{equation}
Only the trace of the tangential stress tensor 
contributes to the change of $H$ under scaling. Locally, this implies that
\begin{equation}
f^a{}_a = - (D+d) {\cal H} + \nabla_a G^a\,,
\end{equation} 
where
$G^a$ is a hypersurface vector field. Modulo a divergence, the
trace is proportional to the integrand. 

For functionals of the form ${\cal H}(g_{ab}, K_{ab})$ one can show that
 $G^a=0$. 
\footnote{\sf To see this, consider the 
response to a deformation of $H$ on a region with boundary:
For functionals of the form ${\cal H}(g_{ab}, K_{ab})$,
Eq.(\ref{eq:delH02}) is replaced by 
\begin{equation}
\delta H = -
 \int dA \, {\bf f}^a \cdot \nabla_a\delta {\bf X}
+ \int dA\nabla_a [ {\cal H}^{ab} {\bf e}_b \cdot \delta\,{\bf n}]\,.
\label{eq:delH03}
\end{equation}
In particular, under a change of scale, $\delta {\bf n}=0$ and
Eq.(\ref{eq:delH03}) reproduces Eq.(\ref{eq:delHs}) without any boundary term.} 
In particular, a scale invariant functional of this
form has vanishing tangential trace: $f^a{}_a=0$. This is {\it
not} true of higher derivative scale invariants: Eq.(\ref{eq:delHS}) does, however,
imply that the trace is the divergence of a hypersurface vector field:
\begin{equation}
f^a{}_a = \nabla_a G^a\,.
\label{eq:trfnabg}
\end{equation}
In particular, in the case of the functional $H_1$ defined
by Eq.(\ref{eq:H1def}), inspection of 
Eq.(\ref{eq:fnabK}) gives for the trace of $f_{ab}$,
\begin{equation} 
f^a{}_a=
{4-D\over 2} (\nabla K)^2
- \nabla^2 K^2/2\,.
\label{eq:trf}
\end{equation}
One identifies $G^a = - \nabla^a K^2/2$. 

\section{\sf Special conformal transformations}

Infinitesimally, a special conformal transformation
induces a change in ${\bf X}$ given by
 \begin{equation}
\delta_{\bf c} {\bf X}= {\bf X}^2\, {\bf c} - 2 ({\bf c}\cdot {\bf X})\, {\bf X}\,.
\end{equation}
The corresponding response of 
the energy is determined using Eq.(\ref{eq:delH02}):
\begin{eqnarray}
\delta_{\bf c} H &=&- \int dA \, {\bf f}^a \cdot \nabla_a\delta_{\bf c} {\bf X}
\nonumber\\
&=& -\int dA \, \left[f^{ab}\, ( {\bf e}_b\cdot\nabla_a\delta {\bf X})
+ f^a \, ({\bf n}\cdot \nabla_a\delta {\bf X})\right]
\nonumber\\
&=&
 2\int dA \,\left[ f^a{}_a \,({\bf c}\cdot {\bf X})
-\, f^a \, {\bf c}\cdot {\bf f}_{0\,a}\right] \nonumber\\
&&
+ 2\int dA f^{ab}
\left[ ({\bf e}_a\cdot {\bf c}) ({\bf e}_b\cdot {\bf X}) -
(a\leftrightarrow b\right)]
\,.
\label{eq:delspc}
\end{eqnarray}
where 
\begin{equation}
{\bf f}_0^a =
({\bf e}^a\cdot {\bf X}) \, {\bf n}
-({\bf n}\cdot {\bf X}) \, {\bf e}^a\,.
\label{eq:f0def}
\end{equation}
The identities 
\begin{equation}
{\bf e}_a \cdot \nabla_b\delta_{\bf c} {\bf X}
= 2\left[ ({\bf e}_a\cdot {\bf c}) ({\bf e}_b\cdot {\bf X}) -
(a\leftrightarrow b\right)]- 2 ({\bf c}\cdot {\bf X})\, g_{ab}\,,
\end{equation}
and
\begin{equation}
{\bf n} \cdot \nabla_a\delta_{\bf c} {\bf X}
= 2\left[ ({\bf e}_a\cdot {\bf X}) ({\bf c}\cdot {\bf n}) -
({\bf e}_a\cdot {\bf c}) ({\bf X}\cdot {\bf n}) \right]
=
2 {\bf c}\cdot {\bf f}_{0\,a}
\end{equation}
have been used on the third line of Eq.(\ref{eq:delspc}). 
In the case of any energy we will consider
$f^{ab}$ is symmetric so that the term
appearing on the last line in Eq.(\ref{eq:delspc}) vanishes.
The equation thus simplifies to 
\begin{equation}
\delta_{\bf c} H
= 2\int dA \,\left[ f^a{}_a \,({\bf c}\cdot {\bf X})
-\, f^a \, {\bf c}\cdot {\bf f}_{0\,a}\right] \,.
\label{eq:delspc0}
\end{equation}
Further simplification is possible using the structure of ${\bf f}^a$.
Using the fact that $f^a = - \nabla_b {\cal H}^{ab}$, 
where ${\cal H}^{ab}$ is  
given by Eq.(\ref{eq:calHdef}),     
the second term appearing on the r.h.s of Eq.(\ref{eq:delspc0}) 
can be cast as   
\begin{equation}
\int dA\, f^a \, {\bf c}\cdot {\bf f}_{0\,a}
=
\int dA\, {\cal H}^{ab} \, {\bf c}\cdot \nabla_b {\bf f}_{0\,a}\,.
\label{eq:idsc1}
\end{equation}
However, the definition of ${\bf f}_0^a$ (\ref{eq:f0def}) gives
\begin{equation}
\nabla_b {\bf f}_{0\,a}= g_{ab}\,{\bf n} + K_{b}{}^c
(({\bf X}\cdot {\bf e}_a)\,
{\bf e}_c - (a\leftrightarrow c))\,.
\end{equation}
The tangential projection is anti-symmetric in $a$ and $b$ and so does not
contribute to the r.h.s of Eq.(\ref{eq:idsc1})
if $f^{ab}= T^{ab}- {\cal H}^{ac} K_c{}^b$ is symmetric. Even when it is not,
it cancels against an identical term appearing on the
last line in Eq.(\ref{eq:delspc}). 
There follows 
the identity 
\begin{equation}
\delta_{\bf c} H =
 2\int dA \,\left[ f^a{}_a \,({\bf c}\cdot {\bf X})
-\, {\cal H}^a{}_a \, ({\bf c}\cdot {\bf n})\right]\,,
\label{eq:delHspc1}
\end{equation}
where ${\cal H}^a{}_a = g_{ab} {\cal H}^{ab}$.
The response of $H$ to an infinitesinal special conformal transformation
has been expressed as a difference
of two terms. Each of therse terms  involves a trace.
Note that in the case of any intrinsic geometrical invariant, 
there exists a representation in which 
the  second term vanishes.

The energy $H$ is conformally invariant 
if and only if 
Eqs.(\ref{eq:trfnabg}) and 
\begin{equation}
f^a{}_a \,({\bf c}\cdot {\bf X})
-\, {\cal H}^a{}_a \, ({\bf c}\cdot {\bf n})=
\nabla_a h^a \,,
\label{eq:cvc}
\end{equation}
are satisfied. $h^a$ is a hypersurface vector field.

We now construct a 
four-dimensional energy, 
polynomial in the curvature and its derivatives, that is consistent with 
these constraints.

\section{\sf Conformally invariants polynomial in  $K_{ab}$}

A scale invariant energy with a
density ${\cal H}$ depending on $g_{ab}$ and $K_{ab}$ 
but not on their derivatives has traceless $f^{ab}$:
$f^a{}_a=0$. To be invariant under
special conformal transformations, one also requires that
\begin{equation}
\int dA\,  {\cal H}^{a}{}_{a}\, {\bf n}= 0\,.
\label{eq:trH0}
\end{equation}
This is clearly satisfied 
if ${\cal H}^{ab}$ is also traceless: ${\cal H}^a{}_a=0$.
It is straightforward to construct polynomial functionals with this property.
These are the well-known invariants involving the 
traceless part of the extrinsic curvature 
tensor $\tilde K_{ab}$ ($\tilde K^a{}_a=0$)
\begin{equation}
\tilde K_{ab}= K_{ab}- {K\over D} g_{ab}  \,.
\end{equation}
Let ${\cal H}$ be a product of terms ${\cal H}_n$,
each of which
is a trace over a 
product of $n$ $\tilde K_{ab}$'s (see definition below Eq.(\ref{eq:bef})); 
${\cal H}_n = {\rm tr}\, \tilde K^n$.
Note that
\begin{equation}
\Pi_{ab}{}^{cd} := 
{\partial \tilde K_{ab}\over \partial K_{cd}} = 
{1\over 2} (\delta_{a}{}^{c} \delta_{b}{}^{d}
+ \delta_a{}^d \delta_b{}^c )  - {1\over D} g_{ab} g^{cd}
\end{equation}
projects out the trace ($\Pi_{ab}{}^{cd} g_{cd}=0$).
For each factor ${\cal H}_n$, we thus find that
\begin{equation} 
g_{ab} {\partial {\cal H}_n\over \partial K_{ab}}=0\,.
\end{equation}
Consequently, ${\cal H}^a{}_a=0$ \cite{Willmore} (see also \cite{conf}).
Two-dimensional surfaces are considered in an appendix.

In four dimensions,  there are two polynomial conformal invariants
constructed this way corresponding to the two independent quartics 
${\rm tr}\, \tilde K^4$ and $({\rm tr}\, \tilde K^2)^2$.
One linear combination of the two is 
the Weyl invariant of the intrinsic geometry \cite{Wald}.

It is possible to satisfy Eq.(\ref{eq:trH0}) in a rather 
less trivial way. It was seen that the translation invariance of any
functional of the form (\ref{eq:Hdef}) implies 
the identity Eq.(\ref{eq:Ef}) between its Euler-Lagrange derivative and the hypersurface 
divergence of 
a stress tensor. If this identity is integrated over a closed hypersurface it follows 
immediately that the Euler-Lagrange derivative satisfies
\begin{equation}
\int dA\, {\cal E}\, {\bf n} =0\,.
\label{eq:calE0}
\end{equation}
Thus, if it is possible to  cast 
${\cal H}^{a}{}_a$ as the Euler-Lagrange derivative of some translationally 
invariant functional 
of the form (\ref{eq:Hdef}),  then Eq.(\ref{eq:trH0}) will be satisfied.
However, one can show that the only  energy densities constructed using $K_{ab}$ 
consistent with this condition are proportional to 
a sum of the symmetric polynomials 
in the principal curvatures. 
Consisitently with scale invariance leaves only the determinant. 
Thus the only conformal invariant generated this way is the
Gauss-Bonnet topological invariant \cite{Spivak}.   

Note that the Paneitz invariant, which has been the 
centre of recent research, is the difference between the Gauss-Bonnet 
and the Weyl invariants \cite{Paneitz,Kiess}. As such, it is an invariant of the intrinsic geometry.

\section{\sf Conformal invariant quadratic in gradients of $K_{ab}$}
\label{sect:quad}

In four dimensions, the functional $H_1$ defined by
Eq.(\ref{eq:nabK2})
is scale invariant but it is not conformally invariant.
It is possible, however, to construct a conformal invariant by 
adding to $H_1$ the integral of an appropriate quartic in $K_{ab}$.
One way to identify this quartic is as follows:

\vskip0.5em \noindent {(i)} First determine how $H_1$ transforms:

\vskip0.5pc
\noindent For $H_1$, neither 
$f^a{}_a$ nor ${\cal H}^a{}_a$ vanishes; one
must contend with the two terms appearing in Eq.(\ref{eq:delHspc1}).  
For $H_1$,  Eq.(\ref{eq:trf}) gives
$f^a{}_a= - \nabla^2 K^2/2$. 
On performing two integrations by parts and using the Gauss-Weingarten equations, one finds
that
\begin{equation}
 \int dA \, f^a{}_a \,({\bf c}\cdot {\bf X})
=
{1\over 2} \int dA \, K^3\, ({\bf c}\cdot {\bf n})\,.
\label{eq:nab1}
\end{equation}
In addition, using
${\cal H}^{ab}=- g^{ab}\nabla^2 K$,
\begin{equation}
\int dA \, {\cal H}^a{}_a  \, ({\bf c}\cdot {\bf n})
=
- 4 \int dA\, \nabla^2 K\,  ({\bf n}\cdot {\bf c})
\,.
\label{eq:nab2}
\end{equation}
The identities  (\ref{eq:nab1}) and (\ref{eq:nab2}) are now substituted into 
Eq.(\ref{eq:delHspc1}) to give for $\delta_{\bf c} H_1 $
\begin{equation}
\delta_{\bf c} H_1 =
 \int dA \, (K^3 + 8\nabla^2 K) \, ({\bf n}\cdot {\bf c})\,.
\label{eq:delcH1}
\end{equation}

\vskip0.5em\noindent {(ii)}
Next note that the r.h.s of Eq.(\ref{eq:delcH1}) can be simplified by using an identity associated with 
the quadratic energy $H_0$
defined by Eq.(\ref{eq:H0}).
Using Eq.(\ref{eq:Ehel}), 
Eq.(\ref{eq:Ef}) implies that
\begin{equation}
 \int dA \, \Big[\nabla^2 K +  ({\rm tr} K^2 - {1\over 2} K^2) K \Big]
\, ({\bf n}\cdot {\bf c})
=0\,,
\label{eq:will}
\end{equation}
for any ${\bf c}$.
This identity allows 
$\delta_{\bf c} H_1$ 
given by Eq.(\ref{eq:delcH1}) to be 
expresssed in terms of a cubic polynomial in $K_{ab}$:
\begin{equation}
\delta_{\bf c} H_1 =
- 8  \int dA \, ({\rm tr}\, K^2 - {5\over 8} K^2) K
\, ({\bf n}\cdot {\bf c})\,.
\end{equation}
Now let ${\cal H}_2= (\nabla K)^2/2+ {\cal H}'$ where
${\cal H}'$ is quartic in $K_{ab}$. If
\begin{equation}
g_{ab}{\partial {\cal H}' \over \partial K_{ab}}
=  8 ({\rm tr}\, K^2 - {5\over 8} K^2) K\,,
\label{eq:trHH}
\end{equation}
then $H=\int dA \,{\cal H}$ will be conformally invariant.
The choice of ${\cal H}'$ is clearly not unique:
however, it is modulo a linear combination of the two conformally covariant quartics,
${\rm tr}\, \tilde K^4$ and $({\rm tr}\, \tilde K^2)^2$.
The simplest choice is a linear combination of 
 the two invariants, $K^4$ and $K^2 {\rm tr}\,  K^2$.
A short calculation  gives
\begin{equation}
{\cal H}'=  K^2\, {\rm tr}\, K^2
- {7\over 16} K^4\,,
\end{equation}
and one identified the following four-dimensional conformally invariant energy:
\begin{equation}
H_2= {1\over 2} \int dA \Big( (\nabla K)^2 - {7\over 8} K^4 +
 2 K^2\, {\rm tr}\, K^2
\Big)
\,.
\label{eq:H2def}
\end{equation}
This identification appears to be new.
Unlike the two quartic conformal invariants in four-dimensions,
the conformal invariance of $H_2$ involves a delicate balance between gradients and  
quartics.

With our choice of quartic ${\cal H}'$, $H_2$
is not positive,
unlike the invariants constructed using ${\rm tr}\, \tilde K^4$ and $({\rm tr}\, \tilde K^2)^2$,
which are. It is, however,  possible to form  
a positive conformally invariant 
gradient energy by 
adding an appropriate linear combination of the 
other invariants. The most general four-dimensional 
conformally invariant energy polynomial in the extrinsic curvature will involve 
a linear combination of all three invariants.

Note that a physically realistic four-dimensional generalization of the 
Willmore energy will involve $H_2$. Consider the Monge description of the 
hypersurface in terms of a height function $h$ above a 
reference plane. With respect to Cartesian coordinates on this plane,
the extrinsic curvature tensor takes the form  
\begin{equation}
K_{ab} = - { \nabla_a\nabla_b h \over (1 + (\nabla h)^2)^{1/2} }\,,
\end{equation}
where now $\nabla_a$ is the flat derivative on this plane.
To lowest order in $h$, 
$K_{ab}\approx - \nabla_a\nabla_b h + {\cal O}(h^3)$.
In the Gaussian approximation, quadratic  in $h$,
all quartics in $K_{ab}$ vanish; in particular, the 
conformal invariants constructed using 
${\rm tr}\, \tilde K^4$ and $({\rm tr}\, \tilde K^2)^2$ both vanish.
For $H_2$ given by Eq.(\ref{eq:H2def}) only the gradient term survives and
one is left with
\begin{equation}
H_2= {1\over 2} \int dA_\perp \, (\nabla \Delta h)^2 + {\cal O}(h^4)\,,
\end{equation}
where $dA_\perp$ is the area element and $\Delta$ is the 
flat Laplacian on this plane. 
The conformal invariance of $H_2$ is, of course, necessarily mutilated in the 
approximation process. 

\section{\sf Generalization}

The construction of the invariant in Section 
(\ref{sect:quad}) suggests that it may be useful to substitute 
Eq.(\ref{eq:trfnabg}) for $f^a{}_a$ into Eq.(\ref{eq:delHspc1}).
If $f^a{}_a $ is  
the Laplacian of some scalar, 
$f^a{}_a = \nabla^2 G$, then one can perform two integrations by parts
to  re-express 
\begin{equation}
 \int dA \, f^a{}_a \,({\bf c}\cdot {\bf X})
=
-  \int dA \, G K\, ({\bf c}\cdot {\bf n})\,.
\label{eq:lapf}
\end{equation}
It is now possible to peel off the space vector in Eq.(\ref{eq:delHspc1})
so that Eq.(\ref{eq:cvc}) is replaced by
\begin{equation}
( G K  
-\, {\cal H}^a{}_a )\, {\bf n} =
\nabla_a {\bf F}^a \,.
\label{eq:cvf}
\end{equation}
A sufficient condition for
conformal invariance is that $f^a{}_a=\nabla^2 G$, and
$G K - {\cal H}^a{}_a $ is an  
Euler-Lagrange derivative. 
${\bf F}^a$ is the corresponding stress tensor.
This is clearly not the only way that 
conformal invariants can occur. In fact,
if the energy depends only on the intrinsic geometry, then ${\cal H}^{ab}$ vanishes
and there are no solutions of this form. An interesting exercise 
would be to identify other energies with an $f^a{}_a$ which is the Laplacian of a scalar.

\section{\sf Discussion}

In the statistical field theory of surfaces,  conformally invariant Hamiltonians 
provide fixed points of the renormalization group flow. 
While the theory of two-dimensional
surfaces is well understood \cite{Nel.Pir.Wei:89},
next to nothing is known about possible four-dimensional counterparts.
The identification of the appropriate invariants is a small first step
towards the formulation of such a  theory.
Before plunging into statistical field theory, however,
there are questions of an elementary nature that should be addressed.
What are the minima of the  conformally invariant energy?
Even without constraints, highly non-trivial vacua appear to be
admitted.  Is there is a useful analogue of Willmore's conjecture \cite{Willmore}? 
The classification of solutions  
will involve topological selection rules beyond the scope of this paper. It would also to know 
if four-dimensional analogues of Goetz and 
Helfrich's egg carton geometries exist \cite{Goetz}.

In the same way that the two-dimensional Willmore functional 
finds an application in relativistic field theory
with the replacement of a Euclidean signature metric by a Lorentzian one, 
it is possible that the four-dimensional
conformally invariant bending energy will find a role in 
 braneworld cosmology \cite{Randall}.
In this context, the generalization of the conformally invariant energy
(\ref{eq:H2def}) to accommodate a curved bulk should be straightforward.

Finally, it should be noted that our construction 
has been framed in the language of classical differential geometry.
Its translation into the language of 
Cartan's exterior differential systems should be straightforward but useful,
especially for addressing issues of a topolgical nature \cite{Bryant}
(see also \cite{Tu}).

\vskip1pc
\noindent{\large \sf Acknowledgments}

\vspace{.5cm}

I thank Riccardo Capovilla,
Markus Deserno and Martin M\"uller for helpful comments.
I am also grateful to Denjoe O' Connor for hospitality at DIAS Dublin where
this work was begun. 
Partial support from CONACyT grant 44974-F is acknowledged.

\appendix

\section{\sf Two-dimensional surfaces}

When $D=2$, the only  
polynomial in $\tilde K_{ab}$,  
with scaling dimension $d=-2$ and vanishing ${\cal H}^a{}_a$
is ${\rm tr}\, \tilde K^2 =
\tilde K^{ab} \tilde K_{ab}$.
The corresponding energy is 
the Willmore energy.

The Gauss-Bonnet topological invariant, 
\begin{equation}
\int dA \, {\rm det} K^a{}_b
\end{equation}
is, of course, also a conformal invariant. For a two-dimensional surface
${\rm det}\, K^a{}_b = {\cal R}/2$.
In fact,  any quadratic invariant in extrinsic curvature is trivially also
conformally invariant. This is because any scalar in the extrinsic curvature
can be expressed as a linear combination of
${\rm tr}\, \tilde K^2=
{\rm tr} \, K^2 - K^2 /D$ and ${\cal R}= K^2 - {\rm tr}\, K^2 $
both of which give rise to conformal invariants when $D=2$.
Thus, in particular, the two quadratics in $K_{ab}$, $K^2$
and ${\rm tr}\, K^2$, also provide conformal invariants. 
This fact can be phrased in an alternative way. 
We note  that both ${\cal H}= K^2$
and ${\cal H}= {\rm tr}\, K^2$, have ${\cal H}^{a}{}_{a} \propto K$
so that Eq.(\ref{eq:trH0}) is satisfied.

\end{document}